\documentclass[aps, prl, 10pt, superscriptaddress, twocolumn]{revtex4-1}

\usepackage{amsmath,amssymb,mathrsfs}
\usepackage{graphicx}

\DeclareMathOperator{\tr}{tr}

\newcommand{\la}{\label}
\newcommand{\bbm}{\begin{multline}}
\newcommand{\eem}{\end{multline}}
\newcommand{\be}{\begin{equation}}
\newcommand{\ee}{\end{equation}}
\newcommand{\bea}{\begin{eqnarray}}
\newcommand{\eea}{\end{eqnarray}}
\newcommand{\p}{\partial}

\newcommand{\cP} {\mathcal{P}}
\newcommand{\cB} {\mathcal{B}}

\newcommand{\comment}[1]{}
\usepackage{color}

\allowdisplaybreaks[1]

\begin{document}

\title{Thermal Hall Effect and Geometry with Torsion.}

\author{Andrey~Gromov}
\affiliation{Department of Physics and Astronomy, Stony Brook University,  Stony Brook, NY 11794, USA}

\author{Alexander G.~Abanov}
\affiliation{Department of Physics and Astronomy and Simons Center for Geometry and Physics,
Stony Brook University,  Stony Brook, NY 11794, USA}

\date{\today}

\begin{abstract}
We formulate a geometric framework that allows to study momentum and energy transport in non-relativistic systems. It amounts to coupling of the non-relativistic system to the Newton-Cartan geometry with \emph{torsion}. The approach generalizes the classic Luttinger's formulation of thermal transport. In particular, we clarify the geometric meaning of the fields conjugated to energy and energy current. These fields describe the geometric background with non-vanishing temporal torsion. We use the developed formalism to construct the equilibrium partition function of a non-relativistic system coupled to the NC geometry in 2+1 dimensions and to derive various thermodynamic relations. 
\end{abstract}


\maketitle


\paragraph{1. Introduction.}

In the seminal work of 1964 Luttinger developed a linear response theory for thermoelectric transport. \cite{luttinger1964theory} An essential part of his approach is the coupling of the many body system to an auxiliary external ``gravitational potential'' conjugated to the energy density. 
The evolution of the energy density is defined by the divergence of energy current, the latter being a fundamental object in the theory of thermal transport. In this paper we identify the appropriate sources of the momentum, energy, and energy current in \emph{non-relativistic} systems. We use the developed general formalism to derive thermodynamic relations involving thermal Hall current in the presence of external fields.

In \emph{relativistic} systems the energy density and the corresponding current are naturally combined into a stress-energy tensor $T^{\mu\nu}$ coupled to an external gravitational field described by the spacetime metric. The energy-momentum and charge conservation laws can be written as 
\be
 \la{RelH}
	\p_\mu T^{\mu\nu} = F^{\nu\rho}J_\rho, \qquad \p_\mu J^\mu=0,
\ee
where $T^{\mu\nu}$ is a stress-energy tensor defined as a response to the external metric $g_{\mu\nu}$. Here, we introduced an electric current $J^\mu$ and an external electromagnetic field $F_{\nu\rho}=\p_\nu A_\rho-\p_\rho A_\nu$. Given a matter action $S$ we can compute the energy-momentum tensor and the electric current as
\be 
 \la{RelTJ}
	T^{\mu\nu} = \frac{2}{\sqrt{g}} \frac{\delta S}{\delta g_{\mu\nu}},  
	\qquad J^\mu = \frac{1}{\sqrt{g}} \frac{\delta S}{\delta A_\mu} \,.
\ee
In the absence of the external sources the first equation in (\ref{RelH}) encodes two conservation laws: conservation of momentum and conservation of energy
\be
	\dot P^j + \p_iT^{ij}=0, \qquad \dot \varepsilon + \p_iJ^i_E = 0,
 \la{NRPEcons}
\ee
where we introduced momentum, energy and energy current as $P^j \equiv T^{0j}$, $\varepsilon = T^{00}$ and $J^i_E = T^{i0}$. These notations will be very natural later on.
In {\it relativistic} systems the stress-energy tensor $T^{\mu\nu}$ (being defined as response to the external space-time metric) is symmetric. This implies equality of momentum and energy current $P^i=J^i_E$. 

In {\i non-relativistic systems} this equality no longer holds. For example, for a single massive non-relativistic particle with mass $m$ moving with velocity $v^i$ we have $P^i = mv^i$ and $J^i_E = \frac{m v^2}{2} v^i$. 

The first result of this letter is the identification of the appropriate sources for the momentum, energy and energy current. We introduce a non-relativistic analogue of (\ref{RelTJ}). This is achieved by replacing the space-time metric $g_{\mu\nu}$ by a different geometric data known as Newton-Cartan (NC) geometry with {\it {torsion}}. We explain how to couple a given non-relativistic system to the NC geometry. The NC geometry has appeared in the context of Quantum Hall effect \cite{son2013newton}, non-relativistic (Lifshitz) Holography \cite{Christensen-2014} and fluid dynamics \cite{Carter-NCfluid}.

While the coupling to NC geometry can be used in any non-relativistic field theory we are mainly motivated by applications to non-relativistic fluid dynamics. In fluid dynamics in addition to standard symmetry constraints of field theory there is an additional set of conditions that ensure that solutions of (\ref{NRPEcons})  are compatible with the (local) second law of thermodynamics \cite{LandauLifshitz-6}. Recently these constraints became a subject of active research in relativistic hydrodynamics \cite{jensen2012towards,Jensen-long}. It turns out that some of these constraints can be obtained systematically demanding that solutions of equations (\ref{RelH}) consistently describe thermal equilibrium in the presence of static external sources \cite{jensen2012towards,banerjee2012constraints}. Here we are interested in non-relativistic applications of these ideas.

The second result of this letter is a construction of the generating functional of Euclidean static correlation functions consistent with local space-time and gauge symmetries. 
Consistency of these static correlation functions with stationary solutions of non-relativistic hydrodynamics provide constraints on the latter.
We note here that equilibrium analysis should be valid for rather general, not necessarily Galilean invariant systems. Throughout the letter we assume that we are in 2+1 dimensions, but most of the analysis is valid in any dimension with obvious modifications.

\paragraph{2. Coupling to Newton-Cartan geometry.} 

Conservation laws (\ref{NRPEcons}) follow from the space and time translation symmetries. Gauging these symmetries will allow us to introduce external fields that naturally couple to momentum, energy and energy current. 

Before going to general formulations we consider an example of free fermions. The action is given by 
\be 
 \la{IQH}
	S = \int dt d^2x \left( i\Psi^\dag \p_0 \Psi - \frac{1}{2m}(\p_A\Psi)^\dag(\p_A\Psi) \right)\,. 
\ee
In order to make this action coordinate independent, {\it{i.e.}} gauge the time and space translations we introduce frame fields (or vielbeins) $E^{\mu}_{a}$ and replace the derivatives in (\ref{IQH}) as follows
\be
 \la{vielbein}
	\p_A \rightarrow E^\mu_A \p_\mu, \quad \p_0 \rightarrow E^\mu_0 \p_\mu\,.
\ee
The second replacement can be understood as a material derivative so that the vielbein $E^\mu_0$ is the velocity field.
Then the action (\ref{IQH}) takes the form
\bea
	S &=& \int dt d^2x e \mathcal{L} \,,
 \nonumber \\
	\mathcal{L} &=&\left(\frac{i}{2}v^\mu(\Psi^\dag  \p_\mu\Psi 
	-  \p_\mu\Psi^\dag \Psi ) -\frac{h^{\mu\nu}}{2m}\p_\mu\Psi^\dag\p_\nu\Psi\right) \,.
 \la{IQH-NC}
\eea
Our conventions $a,b,\ldots=0,1,2$ and $\mu,\nu,\ldots =0,1,2$, also $A,B,\ldots=1,2$ and $i,j,\ldots=1,2$. General coordinate transformations act on the greek indices and local frame transformations act on the latin $a,b,\ldots$ indices.

We have defined a {\it {degenerate}} metric $h^{\mu\nu} = \delta^{AB}E^\mu_AE^\nu_B$, 1-form $n_\mu = e^0_\mu$ and  a vector $v^\mu=E^\mu_0$. Notice, that the spatial part of the metric $h^{ij}$ is a (inverse) metric on a fixed time slice, it is symmetric and invertible. We have denoted its determinant $\det(h^{ij})=h^{-1}$.   
The introduced objects are not independent, but obey the relations 
\be
 \la{NCprop}
	v^\mu n_\mu = 1, \quad h^{\mu\nu}n_\nu = 0.
\ee
These are precisely the conditions satisfied by the NC geometry data \cite{duval2009non,son2013newton}\footnote{It is often convenient to define the ``inverse metric'' \protect{$h_{\mu\nu}=e^A_\mu e^A_\nu$}. It satisfies \protect{$h^{\mu\nu}h_{\nu\rho}=\delta^\mu_\rho -v^\mu n_\rho$ and $h_{\mu\nu} v^\mu=0$} and is fully determined by $v^{\mu},n_{\nu}$ and $h^{ij}$.}. The action (\ref{IQH-NC}) can be viewed as an action (\ref{IQH}) written in an arbitrary coordinate system. The invariant volume element is $dV=edt d^2x$ with $e=\det(e_\mu^a e_\nu^a)$. Due to the spatial isotropy of (\ref{IQH}) the vielbeins naturally combine into the degenerate metric $h^{\mu\nu}$. Similarly, the temporal components of vielbeins (denoted $v^{\mu}$ and $n_{\mu}$) stand aside in (\ref{IQH-NC}) explicitly breaking the (local) Lorentz symmetry down to $SO(2)$. If the physical system was anisotropic the replacement (\ref{vielbein}) would still make sense, but one would have to treat each vielbein as an independent object, {\it i.e.} not constrained by any local symmetries of the tangent space.

In order to couple a generic matter action to the NC geometry one has to proceed in the same way as for the example considered above. Namely, one should modify the derivatives according to (\ref{vielbein}). Then the objects $v^\mu$, $n_\mu$ and $h^{\mu\nu}$  (NC data) will naturally arise (we assume spatial isotropy from now on). When the 1-form $n_\mu$ is not closed we define the Newton-Cartan {\it{temporal torsion}} 2-form as (see Appendix)
\be
\mathcal{T}_{\mu\nu}=\p_\mu n_\nu - \p_\nu n_\mu. 
\ee

In practice, it is convenient to use a particular parametrization of the NC background fields. Let us specify the spatial part $h^{ij}$ of the degenerate metric and assume that $n_\mu = (n_0,n_i)$ and $v^\mu=(v^0,v^i)$ are also specified and satisfy the first relation in (\ref{NCprop}). Then we find from other relations in (\ref{NCprop}) 
$h^{\mu\nu}=\left(\begin{array}{cc}
\frac{n^2}{n_0^2} & -\frac{n^i}{n_0} \\
 -\frac{n^i}{n_0} & h^{ij}\\
\end{array}\right)$, where we defined $n^i=h^{ij}n_j$, $n^2 = n_i n_j h^{ij}$. In this parametrization the invariant volume element is given by $dV =\sqrt{h}n_0dtd^2x$.

The momentum, stress, energy and energy current are identified as responses to the NC geometry as follows
\bea
	P_i &=& \frac{v^0}{\sqrt{h}n_0}  \frac{\delta S}{\delta v^i}, \quad T_{ij} 
	= -\frac{2}{\sqrt{h}n_0}\frac{\delta S}{\delta h^{ij}}\,,
  \la{momentum-stress} \\
	\varepsilon &=& -\frac{1}{\sqrt{h}n_0}\left(n_0\frac{\delta S}{\delta n_0} 
	- v^0\frac{\delta S}{\delta v^0}\right)\,,
 \la{energy} \\
	J^i_E &=& -\frac{1}{\sqrt{h}n_0}\left(n_0\frac{\delta S}{\delta n_i} -
	 v^i \frac{\delta S}{\delta v^0}\right)\,,
 \la{energyJ}
\eea
where we turn off the fields $n_i$ after the variation is taken.

The introduced NC geometry is general and reduces to some cases considered in literature. 
For example, the choice $n_\mu = (1,0,0)$, $v=(1,v^i)$ corresponds to the \emph{torsionless} NC background which turned out to be convenient in studying Galilean invariant actions \cite{2012-HoyosSon,son2013newton,abanov2014,gromov2014density,Haackdiffeo}.

Another particular limit is given by $n_\mu=(e^\psi,0,0)$, $v^\nu=(e^{-\psi},0,0)$. This is an example of the NC geometry with {\it{temporal torsion}}. The torsion is given by 
\be
	\mathcal{T} = e^{\psi}(\p_i\psi) dx^i \wedge dt \,.
\ee
In this case the only non-vanishing component of the torsion tensor is $\mathcal{T}_{0i}$. This NC geometry essentially appeared in the procedure introduced by Luttinger \cite{luttinger1964theory,Halperin-heat}. The field $\psi$ is precisely the ``gravitational potential'' introduced in \cite{luttinger1964theory}. The disadvantage of this choice of geometry is the absence of the field $n_i$ that couples to the energy current. 

In the following we consider a general case keeping all of the components of NC geometry turned on.

\paragraph{3. Examples.}
Let us illustrate how one can derive expressions for conserved currents using the coupling to NC geometry on two examples of physical systems. 

The first example is the system of free fermions. We have already introduced the NC fields into the action of free fermions in (\ref{IQH-NC}). Then the direct application of (\ref{energy}) and (\ref{energyJ}), using equations of motion, and turning off NC fields after the variations we obtain the familiar expressions for energy and energy current in flat space
\bea
	\varepsilon &=& -\frac{1}{2m} (\p_i\Psi)^\dag(\p_i\Psi) \,,
 \\
	J^E_i &=& \frac{i}{4m^2}\left(\p^2\Psi^\dag \p_i\Psi - \p_i\Psi^\dag\p^2\Psi\right) \,.
\eea

The second example is the non-relativistic electrodynamics, {\it{i.e.}} electrodynamics in a medium. The action in the flat background is given by
\be
	S = \int d^2x dt \left(\frac{\epsilon}{8\pi} E^2 - \frac{\mu^{-1}}{8\pi}B^2\right) \,.
 \la{SEB}
\ee
Replacing $\partial_{0}\to v^{\mu}\partial_{\mu}$ and using $h^{\mu\nu}$ instead of contracting spatial indices we obtain from (\ref{SEB})
\be
	S = \int d^2x dt\sqrt{h}n_0\,\, h^{\mu\lambda}\left(\frac{\epsilon}{8\pi}  v^\nu v^\rho  
	- \frac{\mu ^{-1}}{8\pi}h^{\nu\rho}\right)F_{\mu\nu}F_{\lambda\rho} \,,
\ee
where $F_{\mu\nu}=\p_\mu A_\nu-\p_\nu A_\mu$ is the field strength tensor. Again, the direct application of  (\ref{energy}) and (\ref{energyJ}) gives (in flat space)
\bea
	\varepsilon &=& \frac{\epsilon}{8\pi} E^2 + \frac{\mu^{-1}}{8\pi} B^2\,,
 \\
	J^i_E &=& \frac{1}{2\pi} \mu^{-1} \epsilon^{ij}E_jB 
	= \frac{1}{2\pi} {\bf{E}} \times \frac{1}{\mu}{\bf{B}} \,,
 \\
	P^i &=&  \frac{1}{2\pi} \epsilon{\bf{E}} \times {\bf{B}} = \frac{1}{c^2} J^i_E \,.
\eea
One can easily recognize the Poynting vector $J^i_E$ and the momentum density $P^i$ of the electromagnetic field. 

\paragraph{4. Equilibrium.}


We construct a partition function, consistent with time independent, local space and time translations and gauge symmetries. The partition function can be written as a Euclidian functional integral
\be
 \la{W}
	W=-\ln\tr \exp\left\{-\frac{H-\bar\mu N}{\bar T}\right\} = -\ln\int D \Psi D\Psi^\dag e^{-S_E} \,,
\ee
where we introduced a Euclidean action
\be
	S_E[\Psi,\Psi^\dag;A_\mu, n_\mu, v^\mu, h^{ij}] 
	= \int d^2x \sqrt{h}\oint_0^{1/\bar T} d\tau n_0 \mathcal{L}_E \,.
 \la{SE}
\ee
This action is coupled to the NC geometry as explained in the previous section.
The time-independent field $n_0$ can be viewed as an inhomogeneous temperature $T(x)$ defined according to 
\be
	\oint_0^{1/\bar{T}}d\tau n_0\rightarrow \oint_0^{1/T(x)} d\tau^\prime, 
	\quad  \frac{1}{T(x)} = \frac{n_0}{\bar T}\,.
\ee
The NC geometry allows to introduce spatial variations in the size of the compact imaginary time direction.

It is easy to see via usual scaling arguments \cite{banerjee2012constraints} that the Euclidean action has the following functional form
\be
 \la{SEform}
	S_E=S_E\left[\Psi,\Psi^\dag;\frac{A_0}{\bar T},
	\frac{n_0}{\bar T},v^0\bar T,A_i, \frac{n_i}{n_0}\bar T, v^i, h^{ij}\right] \,.
\ee

In (local) equilibrium external fields do not depend on Euclidean time. The generating functional $W$ depends on the temperature $T$ and external sources. We also assume that $W$ can be written as an integral of a local density so that
\be
	W = \int d^2x \sqrt{h}\frac{n_0}{\bar T}\cP\left(\frac{A_0}{\bar T},\frac{n_0}{\bar T},v^0\bar T,A_i, \frac{n_i}{n_0}\bar T, v^i, h^{ij}\right) \,,
 \la{Wgen}
\ee
where we have already replaced the integral over Euclidean time by the overall factor $1/\bar{T}$. 
\paragraph{5. Local time shifts.}
We are mainly interested in the thermal transport, so from now on we set the external field $v^i=0$ and parametrize $v^0=\frac{1}{n_0}\equiv e^{-\psi}$ in order to satisfy (\ref{NCprop}). This field configuration is preserved by the symmetries.

The transformation law of the external field $n_i$ under a local time shift $t \rightarrow t+ \zeta(x) $ takes form
\be\la{timeU1}
\delta(e^{-\psi}n_i) = -\p_i\zeta\,,
\ee
{\it{i.e.}} the field $e^{-\psi}n_i$ transforms like a $U(1)$ gauge field under a local time shift. This field can be regarded as a connection on an $S^1$ bundle over the base manifold, where $S^1$ is the thermal circle. The field strength is related to the NC temporal torsion.

It is convenient to introduce $\mathcal{A}_i = A_i - A_0e^{-\psi}n_i$. This field transform like a gauge field under electro-magnetic gauge transformations and it is {\it{invariant}} under local time shifts.
 
 The symmetry (\ref{timeU1}) implies a local conservation law of the {\it{thermal current}}
 \be
	J^i_Q = -\frac{\bar T}{\sqrt{h}}\left(\frac{\delta W}{\delta e^{-\psi}n_i} 
	+A_0 \frac{\delta W}{\delta A_i} \right) = J^i_E - A_0 J^i \,.
 \ee
 This current is conserved
\be
  \nabla_i J^i_Q=0 \,,
\ee
where $\nabla_i X^i = \frac{1}{\sqrt{h}}\p_i \left(\sqrt{h}\, X^i\right)$ is the covariant divergence.

\paragraph{6. Generating functional in derivative expansion}
We present the partition function as an expansion in derivatives of the external fields. 
We consider the following generating functional
\be 
 \la{W0}
	W = \int d^2x \sqrt{h}\frac{1}{T}\mathcal{P}\left(\mu,T,\mathcal{B},B_E\right) \,,
\ee
where we made the identifications 
\be
 \la{n0-T}
	\frac{1}{T(x)} = \frac{e^{\psi}}{\bar T}, \quad \mu(x) = e^{-\psi}A_0(x)\, ,
\ee
and defined gauge invariant (pseudo) scalars
\be
	\mathcal{B} = \epsilon^{ij}\p_i\mathcal{A}_j,
	\quad B_E =\epsilon^{ij}\p_i (e^{-\psi}n_j)\, .
 \la{BOmega}
\ee
Writing (\ref{W0}) we assumed that both $\mathcal{B}$ and $B$ might be large, while their derivatives are small and can be neglected. We also assumed that gradients of both $\mu$ and $T$ are small.


The generating functional (\ref{W0}) encodes various local thermodynamic quantities and relations. 
For example, the energy (in flat space) can be found with the help of (\ref{energy}), appropriately modified for the presence of the gauge field 
\bea
 	\varepsilon &=& \bar{T}\frac{\delta W}{\delta e^\psi} + T A_0 \frac{\delta W}{\delta A_0}
	=\frac{\p (\cP/T)}{\p (1/T)}-\mu \frac{\p \cP}{\p \mu}
 \nonumber \\
 	&=&\cP + sT + n\mu\, ,
 \la {energy-eq}
\eea
where we made the identifications
\be
	n(x) = \bar{T}\frac{\delta W}{\delta A_0} = -\frac{\p \cP}{\p \mu} 
\ee
and
\be
	s(x) = -\frac{\p \cP}{\p T}.
\ee
The relation (\ref{energy-eq}) suggests that $\cP(\mu,T,\mathcal{B},B_E)$ is the density of the grand thermodynamic potential (in the presence of external fields) and that (\ref{energy-eq}) is the local version of the known thermodynamic relation $\cP = E - \bar TS - \bar\mu N$. 

It is instructive to find the pressure in the presence of external fields, also known as {\it internal} pressure
\be
	P_{int} =\bar T \frac{\delta W}{\delta h^i_{\,\,\, i}} = P_{(0)} - M \mathcal{B} - M_E B_E \,,
 \la{Pint}
\ee
where we have introduced the magnetization $M= e^\psi\frac{\p\cP}{\p \mathcal{B}}$, the energy magnetization $M_E = e^\psi\frac{\p \cP}{\p B_E}$ and $P_{(0)}$ is the pressure at zero magnetic field.

The additional contribution to the pressure given by the second term in (\ref{Pint}) comes from the Lorentz force acting on magnetization currents. The last term of (\ref{Pint}) gives a similar contribution present in non-vanishing background field $B_E$.

\paragraph{7. Magnetization currents.}
While all transport currents vanish in thermal equilibrium, there are still {\it{magnetization}} currents flowing in a material even at equilibrium. These currents cannot be measured in transport experiments \cite{Halperin-heat}. However, e.g., the electric magnetization current can be in principle observed in spectroscopy experiments or by measuring the magnetic field created by moving charges. The energy current can (at least in principle) be observed by the frame drag \cite{2010-RyuMooreLudwig} due to distortions in the gravitational field created by the flow of energy.  In the presence of the inhomogeneous external fields magnetization currents can flow in the bulk of the material, otherwise they are concentrated on the boundary of the sample.

Knowing magnetization currents is important as this knowledge can be used to separate transport currents from the magnetization ones for systems driven out of equilibrium \cite{Halperin-heat}. Also, for a particular case of the chemical potential lying in the excitation gap the magnetization currents are the only currents responsible for the Hall effect \cite{Macdonald-edge}. 

In the following we consider both electric and thermal magnetization currents. They are given, respectively, by
\bea
	J^i &=& \bar T\frac{\delta W}{\delta A_i} = \epsilon^{ij}\p_j M \,,
 \la{elJ} \\
	J^i_Q &=& \epsilon^{ij}\p_j M_E \,.
 \la{enJ}
\eea
The currents (\ref{elJ}) and (\ref{enJ}) are conserved in the presence of arbitrary temperature profile $T(x)$ set by (\ref{n0-T}) and coincide with the ones found in \cite{Halperin-heat,Niu-thermal,ryu-streda} at the level of linear response. 

We note here that usually the energy magnetization $M_E$ is {\it defined} by the Eq. (\ref{enJ}) while the NC "magnetic field" $B_E$ (usually denoted as $B_g$ and referred to as gravimagnetic field) is defined as a quantity thermodynamically conjugated to $M_E$. 
In this work we clarified how one can systematically introduce external fields $n_i$ in {\it non-relativistic} system and  couple the system to $B_E$ (\ref{BOmega}).
Previous approaches explicitly used the presence of Lorentz symmetry \cite{ryu-streda,2010-RyuMooreLudwig} and cannot be applied in majority of condensed matter systems. 

\paragraph{8. Streda formulas.}
It is possible to the express Hall conductivity and other parity odd responses purely in terms of derivatives of thermodynamic quantities. We define electric and thermal conductivities as
\bea
	J^i &=& \epsilon^{ij} \left(\sigma_H \p_i \mu + \sigma_H^T \p_i T \right) \,,
 \\
	J_E^i &=& \epsilon^{ij} \left(\kappa_H^\mu \p_i \mu + \kappa_H \p_i T \right) \,.
\eea
Comparing with (\ref{elJ}-\ref{enJ}) we obtain using the Maxwell's relations \footnote{As $d \cP =-sdT-nd\mu-Md\mathcal{B}-M_Ed\Omega$ we have $\p M/\p \mu = \p n/\p B$ etc.}
\bea
 	\sigma_H &=& \left(\frac{\p M}{\p \mu}\right)_{T,\cB,B_E} 
	=\left(\frac{\p n}{\p \mathcal B}\right)_{T,\mu,B_E}\, ,
  \la{sigma1} \\ 
	\sigma^T_H &=& \left(\frac{\p M}{\p T}\right)_{\mu,\cB,B_E} 
	=\left(\frac{\p s}{\p \mathcal B}\right)_{T,\mu,B_E}\, ,
 \la{sigma2} \\ 
  	\kappa^\mu_H &=& \left(\frac{\p M_E}{\p \mu}\right)_{T,\cB,B_E} 
	=\left(\frac{\p n}{\p B_E}\right)_{T,\mu,\cB}\, ,
  \la{kappa1} \\
   	\kappa_H &=&\left( \frac{\p M_E}{\p T}\right)_{\mu,\cB,B_E} 
	=\left(\frac{\p s}{\p  B_E}\right)_{T,\mu,\cB}\,.
 \la{kappa2}
\eea
These are thermodynamic Streda-type formulas \cite{Streda:1982qf,Streda:1983fv} for the response coefficients.
 
\paragraph{9. Galilean and Lorentz symmetries.} 

So far we assumed that the (un-perturbed) system under consideration is gauge invariant, spatially isotropic and homogeneous, and time translation invariant. In this general case there are no additional relations between electric current, momentum and energy current. Several new relations appear if additional symmetries are present. For simplicity, we assume below that the underlying microscopic system consists of charged particles of a single species or several species with the same $e/m$ ratio. 

If the system is {\it Galilean} invariant the electric current is proportional to the momentum $J^i=\frac{e}{m}P^i$, therefore, the magnetization density is proportional to the density of the angular momentum $M = \frac{e}{m}L_z$.
Then from (\ref{sigma1}) we have
\be
	\sigma_H = \frac{e}{m}\left(\frac{\p L_z}{\p  \mu}\right)_{T,\cB,B_E} \,,
 \la{streda1}
\ee
that is Hall conductivity can be expressed in terms of derivatives of the angular momentum.

If the system is {\it Lorentz} invariant then there is an additional equality between momentum and energy current as we pointed out in the introduction $J^i_E = P^i$ and, therefore, $M_E = L_z$. Therefore, we have another version of Streda formula for thermal Hall conductivity \cite{ryu-streda}
\be
	\kappa_H = \left(\frac{\p L_z}{\p  T}\right)_{\mu, \cB,B_E} \,.
 \la{streda2}
\ee


In general case, when no additional symmetries are present the angular momentum is {\it not} related to either electric or thermal magnetization and the relations (\ref{streda1})-(\ref{streda2}) do not hold.


\paragraph{10. Conclusions.}
To conclude, it is shown that coupling the physical system to the Newton-Cartan geometry introduces the appropriate sources for energy, momentum, and energy current. Variations of the action with respect to different components of the NC geometry give familiar expressions for energy, momentum, and energy current densities. It turns out that in order to introduce the temperature gradients one has to couple a physical system to the NC geometry with {\it{temporal torsion}}.

The developed formalism was used to construct a general equilibrium partition function of a gapped non-relativistic system. With the partition function at hand known thermodynamic relations  have been obtained in the presence of external gauge and Newton-Cartan fields. It was found that upon linearization the found general expressions for electric and thermal magnetization currents agree with the linear response expressions known in literature.

The constructed formalism is expected to have many potential applications in condensed matter systems and hydrodynamics. For example, the general {\it geometric} effective action constructed in the presence of the torsional NC background will not be restricted by the the Lorentz symmetry and, therefore, is more natural in condensed matter context. The Galilean symmetry can be implemented by adding additional constraints on the action coupled to NC geometry data. The generalization to systems with internal degrees of freedom such as spin may prove to be of interest in the context of spin Hall effect.

We acknowledge discussions with B. Bradlyn, A. Cappelli, G. Monteiro, S. Moroz, M. Rocek, D. Son and especially K. Jensen.
The work of A.G.A. was supported by the NSF under grant no. DMR-1206790. 

During the preparation of this work we were made aware of complementary results by Bradlyn and Read \cite{bradlyn-heat}.

After the work was completed we learned about the work \cite{son-torsion} where the NC geometry with torsion was related to the energy transport in non-relativistic systems.

\section{Appendix}
\paragraph{Construction of the NC geometry.}
In this appendix we review the relation of NC geometry data with the familiar Einstein-Cartan (EC) geometry (also known as first order formalism or triad formalism). The NC geometry can be understood as a generalization of the latter for the cases where Lorentz symmetry is absent.

The geometric data of EC geometry consists of four objects: driebeins $e^a_\mu$ and their inverse $E^\mu_a$, spin connection $\omega_{\mu\,\,\,\,b}^{\,\,a}$ and torsion $T^a_{\mu\nu}$ \cite{weinberg-book}. 

Driebens satisfy the following relations
\be 
 \la{prop}
	g_{\mu\nu} = \eta_{ab}e^a_\mu e^b_\nu, \quad g^{\mu\nu} 
	=\eta^{ab} E^\mu_aE^\nu_b, \quad \delta^\mu_\nu 
	= \delta^a_b E^\mu_a e^b_\nu \,,
\ee
where $g_{\mu\nu}$ is space-time metric and $\eta_{ab}$ is a flat metric in tangent space. The geometric data satisfies the Cartan structure equations.
\be
 \la{Cartan}
	de^a + \omega^a_{\,\,\,b} \wedge e^b = T^a \,.
\ee
We impose constraints on these equations and obtain the essential ingredients of NC geometry. 

First, we split (\ref{Cartan}) into temporal and spatial parts and impose the {\it{galilean}} constraint 
\be
 \la{GalC}
	\omega^A_{\,\,\,\,0} = 0 \,.
\ee
This constraint has a simple physical meaning: the part of the spin connection, responsible for Lorentz boosts vanishes identically.

In order to simplify the discussion we also impose $\omega^0_{\,\,\,A} =0$ and $T^A =0$. Then (\ref{Cartan}) takes form
\be
 \la{NC}
	de^0 = T^0\equiv\mathcal{T}, \qquad de^A + \omega^A_{\,\,\,B}\wedge e^B = 0 \,.
\ee
Notice, that while these equations are still covariant in space-time, the tangent space has lost the Lorentz symmetry. From the objects that appear in (\ref{NC}) together with relations (\ref{prop}) we can construct all of the NC geometry data. In particular, the Eq.~(\ref{NC}) clarifies why we refer to $\mathcal{T}$ as temporal torsion.

%
%
%
%

\bibliographystyle{my-refs}




\bibliography{abanov-bibliography}

\end{document}